\documentclass[aps,pra,epsfigure,twocolumn,superscriptaddress]{revtex4-2}
\usepackage[colorlinks=true,linkcolor=blue,urlcolor=blue,citecolor=blue,pdfusetitle]{hyperref}
\usepackage[utf8]{inputenc}
\usepackage[english]{babel}
\usepackage{mathtools}
\usepackage{amsmath}
\usepackage{amssymb}
\usepackage{amsfonts}
\usepackage{enumerate}
\usepackage{bbold}
\usepackage{bbm}

\usepackage[caption = false]{subfig}
\usepackage{graphicx}

\usepackage{latexsym}
\usepackage{physics}
\usepackage{braket}

\usepackage{txfonts}

\usepackage[cal=boondox]{mathalpha}
\renewcommand{\bm}[1]{\boldsymbol{#1}}

\renewcommand{\ketbra}[2]{\ket{#1}\!\bra{#2}}

\DeclareFontFamily{U}{cmsy}{\skewchar\font48 }
\DeclareFontShape{U}{cmsy}{m}{n}{%
	<-5.5>\mathalfa@calscaled cmsy5%
	<5.5-6.5>\mathalfa@calscaled cmsy6%
	<6.5-7.5>\mathalfa@calscaled cmsy7%
	<7.5-8.5>\mathalfa@calscaled cmsy8%
	<8.5-9.5>\mathalfa@calscaled cmsy9%
	<9.5->\mathalfa@calscaled cmsy10}{}
\DeclareMathAlphabet{\mcal}{U}{cmsy}{m}{n}

\newcommand{\intn}{\int\!}

\usepackage{orcidlink}

\begin{document}


\title{Deviations from thermal light statistics in ensembles of independent two-level emitters}

\author{M. Bojer}
\email{manuelbojer6@gmail.com}
\affiliation{Friedrich-Alexander-Universität Erlangen-Nürnberg, Quantum Optics and Quantum Information, Staudtstr. 1, 91058 Erlangen, Germany}
\author{A. Cidrim}
\affiliation{Departamento de F\'isica, Universidade Federal de S\~ao Carlos,
	Rodovia Washington Lu\'is, km 235—SP-310, 13565-905 S\~ao Carlos, SP, Brazil}
\author{R. Bachelard}
\affiliation{Departamento de F\'isica, Universidade Federal de S\~ao Carlos,
	Rodovia Washington Lu\'is, km 235—SP-310, 13565-905 S\~ao Carlos, SP, Brazil}
\author{J. von Zanthier}
\affiliation{Friedrich-Alexander-Universität Erlangen-Nürnberg, Quantum Optics and Quantum Information, Staudtstr. 1, 91058 Erlangen, Germany}

\date{\today}

\begin{abstract}
We investigate the light statistics of an ensemble of independent motionless two-level atoms in a product state. We identify the conditions under which the cold atomic ensemble emits thermal light statistics characterized by the Gaussian Moment Theorem. For the theorem to hold, we  derive for each correlation order two conditions on the atom number and the ratio of coherent to incoherent light emission. We further discuss their validity for atoms either in a pure or mixed state. Our results contribute to the understanding of the generation of thermal light by two-level atoms without interactions among the emitters. 
\end{abstract}

\maketitle

\section{Introduction}
\label{sec:Introduction}

Light sources are characterized not only by their mean power, but rather by the full statistics of the field they radiate. These statistics are described by correlation functions of arbitrary orders, which describe the correlations between the electric field at different points of observation and different times, as introduced by Glauber in 1963~\cite{Glauber1963Coh}. Yet, while the most general treatment addresses the electric field and the detection via quantum mechanical operators, the light statistics can also be assessed using a classical approach, characterized by classical correlations of all field terms~\cite{Loudon:book, gerry2005introductory}. Within the classical approach, the correlation functions can be shown to obey several inequalities, which set a boundary for "classical light" -- such that any field violating such an inequality can be considered as quantum. Antibunching, which reflects an increasing probability with time of emitting a second photon after a first one, is a key example of "quantum light"~\cite{Kimble1977}. It is today sought after for quantum cryptography~\cite{Beveratos2002} and other quantum technology protocols, but it is also a signature of the quantized structure of energy levels of atoms.

Despite the quantum nature of all matter scattering light, most of the light surrounding us is "thermal", i.e., its field exhibits a classical probability distribution of a zero-mean Gaussian function. 
In particular, all the moments of the distribution of this field are determined by the so-called Gaussian Moment Theorem (GMT)~\cite{mandel1995optical}, also known as Isserlis' theorem~\cite{Isserlis1918} or Wick's probability theorem~\cite{Wick1950}. It states that all higher-order moments of the light field can be expressed by sums and products of second-order moments. This situation typically occurs due to the very large number of emitters of macroscopic sources as well as the existence of phase-randomization mechanisms to ensure the phase fluctuations~\cite{Loudon:book, Lassegues2023}.

On the other side, the absence of such decoherence mechanism can lead to peculiar light statistics~\cite{Wolf2020, Bojer2025,Singh2025}. Indeed, even for a single emitter the light can be decomposed as the sum of coherent (elastically scattered) and incoherent (spontaneously emitted) components~\cite{LopezCarreno2018, ZubizarretaCasalengua2020, Phillips2020, Hanschke2020}. In the case of multiple (independent) emitters, the former produces an interference pattern, which in turn leads to a spatial modulation of the light statistics, as recently demonstrated for two trapped and laser-cooled ions \cite{Wolf2020} as well as in trapped and laser-cooled mesoscopic ion chains~\cite{Singh2025}. The convergence to thermal light statistics in ensembles of independent motionless emitters is thus an open question, which we address in this paper.

More specifically, in this manuscript we investigate the conditions under which an ensemble of independent motionless two-level atoms produces light similar to that of a classical thermal light source, that is, the regime in which it obeys the GMT. We derive two conditions for this to occur, involving as critical parameters the number of atoms  as well as the ratio of coherently to incoherently scattered light emitted by the atoms. The derivation of leading order corrections due to finite-size effects and spin coherence provides a guide on the deviation from thermal statistics to be expected in cold atomic ensembles.

The paper is organized as follows: In Sec.~\ref{sec:Thermal_states_and_GMT}, we introduce thermal light sources and the Gaussian Moment Theorem. Then, in Sec.~\ref{sec:Conditions_for_GMT_m_equal_n} and~\ref{sec:Conditions_for_GMT_m_unequal_n}, we present the atomic system under consideration, state the above mentioned two conditions, and exemplify and validate our conditions on different examples of atomic systems. Afterwards, in Sec.~\ref{sec:Comparison}, we compare our findings for two-level atoms with those of classical emitters. Finally, we summarize our results and conclude in Sec.~\ref{sec:Conclusion}.

\section{Thermal states and Gaussian Moment Theorem}
\label{sec:Thermal_states_and_GMT}

Let us first introduce thermal states: their field are complex Gaussian variables which satisfy the Gaussian momentum theorem. Consider a set of $M$ modes of the electromagnetic field described by the Hamiltonian $\hat{H}_M=\sum_{j=1}^M\hbar \omega_j \hat{a}_j^\dagger \hat{a}_j$, with $\omega_j$ the frequency of mode $j$ and $\hat{a}_j$ ($\hat{a}_j^\dagger$) its annihilation (creation) operator. For an inverse temperature $\beta$, the associated thermal state is given by
\begin{subequations}
\begin{align}
    &\hat{\rho} = \frac{e^{-\beta \hat{H}_M}}{\Tr[e^{-\beta \hat{H}_M}]} = \intn d^{2M}\alpha_j P(\alpha_1,...,\alpha_M) \ketbra{\alpha_j}{\alpha_j},
    \\ &P(\alpha_1,...,\alpha_M) = \prod_{j=1}^M \frac{1}{\pi\braket{\hat{a}^\dagger_j \hat{a}_j}}e^{-\frac{|\alpha_j|^2}{\braket{\hat{a}^\dagger_j \hat{a}_j}}},
\end{align}
\end{subequations}
with $\ket{\alpha_j}$ the coherent states of amplitude $\alpha_j$ for mode $j$. $P$ is the Glauber-Sudarshan $P$ function for $M$ \textit{independent} modes, so that the total density operator is the tensor product of the individual ones. Hereafter, the expectation value of a normally ordered operator $\hat{X}(\hat{a}_1,...,\hat{a}_n,\hat{a}^\dagger_1,...,\hat{a}^\dagger_m)$ can be calculated by
\begin{align}
    &\braket{\hat{X}(\hat{a}_1,...,\hat{a}_n,\hat{a}^\dagger_1,...,\hat{a}^\dagger_m)} = \mathrm{Tr}\left[\hat{\rho} \hat{X}(\hat{a}_1,...,\hat{a}_n,\hat{a}^\dagger_1,...,\hat{a}^\dagger_m)\right]\nonumber\\
    &= \intn d^{2M} \alpha_j P(\alpha_1,...,\alpha_M) \hat{X}(\alpha_1,...,\alpha_n,\alpha^*_1,...,\alpha^*_m)\,,
\end{align}
i.e., evaluating a classical expectation value with respect to the quasi probability distribution $P$, which corresponds to the usual optical equivalence theorem.

For independent thermal states, the joint $P$ function is a  multivariate zero-mean complex Gaussian function, whereby the GMT provides a relation between the higher-order moments of the complex variates $\alpha_1,...,\alpha_M$ with the second-order ones. Let us introduce the sets $I = \{i_1,...,i_m\}$ and $J = \{j_1,...,j_n\}$ of $m$ and $n$ indices, respectively. If $m=n$, we define  $P_{\sigma} = \{\{i_1, j_{\sigma(1)}\},...,\{i_m, j_{\sigma(m)}\}\}$ a pair partition of the multiset $I \cup J$, associated with a permutation $\sigma \in S_m$, where $S_m$ is the set of permutations of $m$ elements. Then the GMT for zero-mean Gaussian complex variates states that~\cite{mandel1995optical}
\begin{align}
    \braket{\alpha_{i_1}^*...\alpha_{i_m}^*\alpha_{j_n}...\alpha_{j_1}}=
	\begin{cases}
            \sum\limits_{\sigma \in S_m}\prod\limits_{\{i,j\}\in P_{\sigma}}\braket{\alpha_{i}^*\alpha_{j}} & \textrm{if } m=n\,,\\
		0 & \textrm{if }  m\neq n\,.
	\end{cases}
\end{align} 

In the context of quantum optics, introducing the positive (negative) electric field operator at position $\bm{r}$, $\hat{E}^{+}(\bm{r})$ ($\hat{E}^{-}(\bm{r})$), and the field correlation functions~\cite{Glauber1963Coh}
\begin{equation}
    g^{(m,n)}(\bm{r}_1,...,\bm{r}_{m+n}) = \frac{\langle \hat{E}^{-}(\bm{r}_1)...\hat{E}^{-}(\bm{r}_m)\hat{E}^{+}(\bm{r}_{m+1})...\hat{E}^{+}(\bm{r}_{m+n})\rangle}{\sqrt{\braket{\hat{E}^{-}(\bm{r}_1)\hat{E}^{+}(\bm{r}_{1})}...\braket{\hat{E}^{-}(\bm{r}_{m+n})\hat{E}^{+}(\bm{r}_{m+n})}}},\nonumber
\end{equation}
the GMT  translates into
\begin{align}
\label{eq:GMT_Corr}
	&g^{(m,n)}(\bm{r}_1,...,\bm{r}_{m+n}) = 
    \begin{cases}
            \sum\limits_{\sigma \in S_m} \prod\limits_{\{i,j\}\in P_{\sigma}} g^{(1,1)}(\bm{r}_i,\bm{r}_{j}) & \textrm{if } m=n\,,\\
		0 & \textrm{if } m\neq n\,,
	\end{cases}
\end{align}
where the two sets are given by $I = \{1,...,m\}$ and $J = \{m+1,...,m+n\}$. Hence, provided the electric field is a Gaussian variable, the GMT introduces a relation between field correlation functions of different orders. Note that while the present work focuses on equal-time correlations, it can straightforwardly be generalized to multiple-time correlation functions by substituting $\bm{r}_j$ by ($\bm{r}_j,t_j$), with $t_j$ the time at which the field is observed at position $\bm{r}_j$. For clarity, in the case $m=n$, the correlation function will be noted $g^{(m)}$. Furthermore, for equal points of observation ($\bm{r}_i=\bm{r}_{j}=\bm{r}$) the function is referred to as field auto-correlation function and noted $g^{(m)}(\bm{r})$.

\section{Conditions for the GMT in two-level systems for $m=n$\label{sec:Conditions_for_GMT_m_equal_n}}

Let us now consider a set of $N$ two-level atoms at fixed random positions $\bm{R}_\mu$, with ground state $\ket{g}$ and excited state $\ket{e}$. We assume that the atoms are all in the same state, with negligible interactions between them, so that the system state can be described by a tensor product $\hat{\rho} = \otimes_{\mu=1}^{N} \hat{\rho}_{\mu}$ of the same single-atom state $\hat{\rho}_{\mu}$. 
Then, neglecting the dipole radiation pattern, the positive electric source field operator reads $\hat{E}^{+}(\bm{k}) = \sum_{\mu=1}^{N} e^{-i \bm{k}. \bm{R}_\mu} \hat{\sigma}^-_\mu$
with $k_0= 2 \pi/\lambda$ the wavenumber of the atomic transition at $\lambda$, $\bm{k}=k_0 \hat{\bm{r}}$ and $\hat{\sigma}^-_\mu$ ($\hat{\sigma}^+_\mu$) the lowering (raising) operator of the $\mu$th atom~\cite{agarwal1974quantum}. Note that if the atoms are driven by a near-resonant plane wave laser with wave vector $\bm{k}_L$, the resulting phase on the coherences can be absorbed by redefining $\bm{k}\rightarrow \bm{k}-\bm{k}_L$, maintaining a unique $\hat{\rho}_{\mu}$.

\subsection{Conditions for the GMT}
\label{sec:Cond_for_GMT_m_equal_n}
For this atomic ensemble, we derive in Appendix~\ref{sec:AppA} two conditions which need to be fulfilled to observe thermal light statistics for $m=n$. The first one, which we call \textit{finite-$N$ condition}, restricts the correlation order $m$ up to which thermal light statistics are observed for a certain number of atoms $N$:
\begin{align}
\label{eq:finite_N_cond}
    \frac{m! m (m-1)}{2N} \ll 1\,.
\end{align}
We note that $g^{(m)}=0$ for $m> N$ since $m>N$ simultaneous photons cannot be detected from a cloud of $N$ two-level emitters. This shows that true gaussian statistics, that is, at arbitrary order, can only be fulfilled in the limit $N\to\infty$.

The second condition, which we call \textit{spin-coherence condition}, restricts the single-atom coherence $\braket{\hat{\sigma}^\pm}$ with respect to the fluctuation $\delta\hat{\sigma}^\pm = \hat{\sigma}^\pm - \braket{\hat{\sigma}^\pm}$. For $m=n$, the condition reads
\begin{align}
\label{eq:spin_coh_cond_1}
R^2 = \left(\frac{\braket{\hat{\sigma}^+} \braket{\hat{\sigma}^-}}{\braket{\delta\hat{\sigma}^+ \delta\hat{\sigma}^-}}\right)^2 \ll \frac{4}{N^2 m (m-1)}\,, 
\end{align}
where $R$ quantifies the ratio of coherent to incoherent (spontaneously emitted) radiation of a single atom.  Eq.~\eqref{eq:spin_coh_cond_1} sets an upper bound on $R$ so that the correlation function $g^{(m)}$ obeys the GMT. Indeed, it has been shown that the presence of coherent radiation in large atomic systems can lead to extreme values for the light statistics, both in terms of anti- and superbunching~\cite{Singh2025,Bojer2025}.

Let us emphasize that both conditions are upper bounds, i.e., they represent the strictest conditions. They can be relaxed in two ways. First, depending on the directions of observation ($\bm{k}_1,...,\bm{k}_{2m}$), the factor $m!$ in the finite-$N$ condition can be neglected. For example, for the direction $\bm{k}=\bm{0}$, which corresponds to an observation in the direction of the driving laser, we obtain the condition $m (m-1)/2N \ll 1$ instead of Eq.~\eqref{eq:finite_N_cond}. Second, the spin-coherence condition is derived using $\bm{k}=\bm{0}$ for all $\bm{k}$. This corresponds to maximally constructive interference, so the light intensity scales as $N^2$. In a different direction than $\bm{k}=\bm{0}$, the intensity may be much weaker and the $N$-dependence of the spin-coherence condition may be loosened (see examples below). In particular, if one considers a disordered system of atoms, where the atomic motion leads effectively to an average over realizations, one may rather consider the average intensity, which scales as $N$ according to speckle theory~\cite{Goodman2020}, rather than $N^2$.

\subsection{Deviations from the GMT}
\label{sec:Dev_from_GMT_m_equal_n}

Let us now discuss the deviations which arise from the violation of conditions~\eqref{eq:finite_N_cond} and~\eqref{eq:spin_coh_cond_1} in the case of a cloud of independent motionless two-level atoms, corresponding to signatures of non-Gaussianity of the field. We consider a cloud either coherently excited with a short pulse and thus in a coherent superposition state, or driven to steady state with a cw laser. The deviations from Gaussian statistics, hereafter denoted $\delta g^{(m)}(\bm{k}_1,...,\bm{k}_{2m})$, can be written as:
\begin{widetext}
\begin{align}
\delta g^{(m)}(\bm{k}_1,...,\bm{k}_{2m}) \coloneqq \sum\limits_{\sigma \in S_m} \prod\limits_{\{i,j\}\in P_{\sigma}} g^{(1)}(\bm{k}_i,\bm{k}_{j}) - g^{(m)}(\bm{k}_1,...,\bm{k}_{2m})=\delta g^{(m)}_{N}(\bm{k}_1,...,\bm{k}_{2m}) + \delta g^{(m)}_{\mathrm{coh}}(\bm{k}_1,...,\bm{k}_{2m})\,,
\end{align}
\end{widetext}
where $\delta g^{(m)}_{N}(\bm{k}_1,...,\bm{k}_{2m}) = \delta g^{(m)}(\bm{k}_1,...,\bm{k}_{2m}){\big|_{ \braket{\hat{\sigma}^-}=0} }$
describes the deviation due to the finite number $N$ of emitters, and $\delta g^{(m)}_{\mathrm{coh}}(\bm{k}_1,...,\bm{k}_{2m})$ the deviation due to the finite spin coherence.

\subsubsection{Coherent spin states}
\label{sec:Pulse_excited}

When excited with a coherent short pulse, the atomic system is found in a coherent superposition state $\ket{\psi}=\cos(\theta/2)\ket{g}-i\sin(\theta/2)\ket{e}$, with $\theta$ the pulse area. The associated population is $\braket{\hat{\sigma}^+\hat{\sigma}^-} = \sin^2(\theta/2)$, the coherence $\braket{\hat{\sigma}^\pm} =\mp i\sin(\theta/2)\cos(\theta/2)$, and the fluctuation $\braket{\delta\hat{\sigma}^+ \delta\hat{\sigma}^-} = \sin^4(\theta/2)$.
In the particular case of a fully inverted ensemble, which corresponds to $\theta=\pi$ and $\ket{\psi}=\ket{e}$, there is no spin coherence, so that the associated condition~\eqref{eq:spin_coh_cond_1} is trivially fulfilled. In this case, since $\delta g^{(m)}_{\mathrm{coh}} = 0$, only the finite-$N$ deviation $\delta g^{(m)}_{N}$ remains. This leads, up to third order of the correlation functions, to:
\begin{widetext}
\begin{align}
    \delta g^{(2)}_{N}(\bm{k}_1,...,\bm{k}_4) &= -\frac{2}{N^2} S(\bm{k}_1+\bm{k}_2-\bm{k}_3-\bm{k}_4)\\
    \delta g^{(3)}_{N}(\bm{k}_1,...,\bm{k}_6) &= \underbrace{-\frac{1}{2N^3}\sum_{\sigma,\sigma' \in S_3} S(\bm{k}_{\sigma(1)}-\bm{k}_{3+\sigma'(1)})S(\bm{k}_{\sigma(2)}+\bm{k}_{\sigma(3)}-\bm{k}_{3+\sigma'(2)}-\bm{k}_{3+\sigma'(3)})}_{\text{first-order deviation}} \underbrace{+\vphantom{\sum_{\sigma,\sigma' \in S_3}} \frac{12}{N^3} S(\bm{k}_1+\bm{k}_2+\bm{k}_3-\bm{k}_4-\bm{k}_5-\bm{k}_6)}_{\text{second-order deviation}}\,,\nonumber
\end{align}    
\end{widetext}
where we have introduced the structure factor $S(\bm{k}) = \sum_{\mu=1}^N e^{i \bm{k}\cdot\bm{R}_{\mu}}$. Since $|S(\bm{k})|\leq N$, the deviation $\delta g_{N}^{(2)}$ is at most $\frac{2}{N}$. This is consistent with the result for the intensity autocorrelation (i.e., for  $\bm{k}_1=\bm{k}_2=\bm{k}_3=\bm{k}_4$): $N$ spontaneously emitting two-level atoms exhibit $g^{(2)}(\bm{r})=2-2/N$, while $g^{(2)}(\bm{r})=2$ is expected for Gaussian statistics. Thereby, the factor of $2$ denotes the number of permutations $m!$, which needs to be multiplied by the coefficient given in Eq.~\eqref{eq:class_non_class_coeff} in App.~\ref{sec:AppA} to obtain the full deviation. In the case of $g^{(3)}$, the first-order deviation is at most $18/N$, whereas the second-order deviation is at most $12/N^2$.



\begin{figure}[t!]
\centering
\includegraphics[width=.5\textwidth]{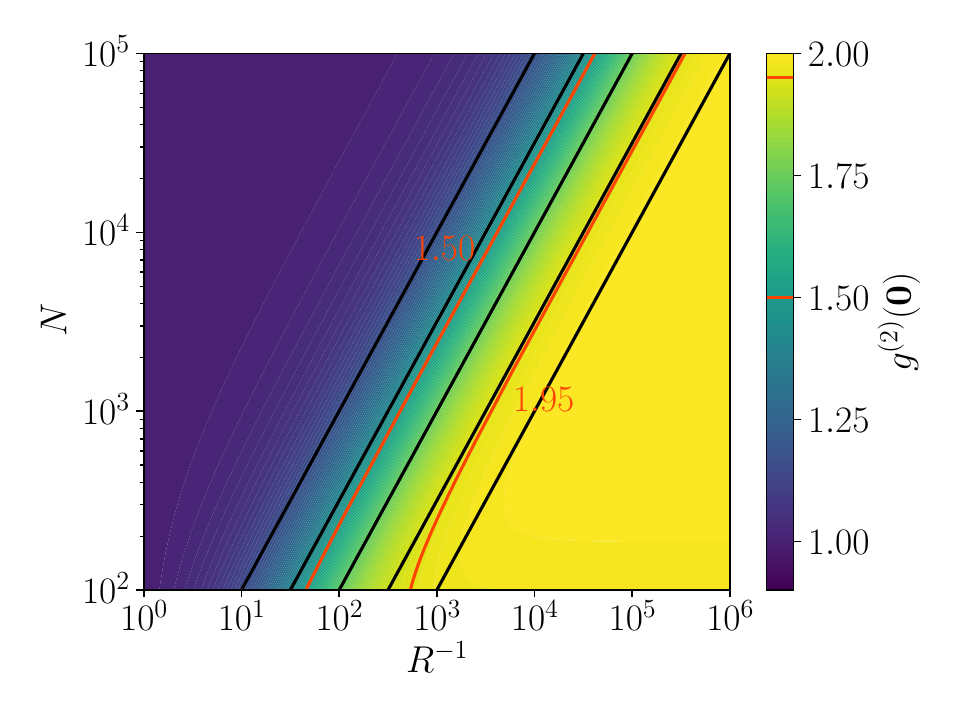}
\caption{Second-order autocorrelation function $g^{(2)}(\bm{0})$ as a function of $N$ and of the inverse ratio $R^{-1}=\tan^2(\theta/2)$. Decreasing the ratio of the coherences and fluctuations for fixed $N$ leads to a transition towards the Gaussian Moment Theorem characterized by a value of $g^{(2)}(\bm{0})=2$. In red we highlight two contour lines of $g^{(2)}(\bm{0})$ at $1.50$ and $1.95$. By comparison with the contour lines of $(N R)^2=[N\cot^2(\theta/2)]^2$ (black solid lines) the quadratic behavior of the spin-coherence deviation is clearly visible, except for smaller $N$ and values of $g^{(2)}(\bm{0})$ close to $2$, where the finite-$N$ deviation becomes relevant (see red $1.95$ curve).
}
\label{Fig:g2_map_coh}
\end{figure}

Let us now discuss the deviations from the GMT due to  spin-coherence, by considering atoms excited by a pulse with $\theta<\pi$. We first note that the finite-$N$ corrections $\delta g^{(m)}_{N}$ are the same as for the fully inverted ensemble, since these are state independent. We further see that in the present case the ratio of coherently to incoherently scattered power is $R=\cot^2(\theta/2)$. 
Since the coherent contribution is largest in the forward direction, we here focus on the autocorrelations with $\bm{k}_j=\bm{0}\,\forall j$, which we denote by $|\delta g^{(m)}_{\mathrm{coh}}(\bm{0})|$ for brevity. Then the deviation from Gaussian statistics scales, at leading order, as (see App.~\ref{sec:AppA})
\begin{equation}
    \delta g^{(m)}_{\mathrm{coh}}(\bm{0})\propto (NR)^2\,.
\end{equation}
In Fig.~\ref{Fig:g2_map_coh}, we show the second-order autocorrelation function $g^{(2)}(\bm{0})$ against $N$ and the inverse ratio $R^{-1}=\tan^2(\theta/2)$, where for a given number of atoms a transition towards the GMT, i.e., a value of $2$ of $g^{(2)}(\bm{0})$, is visible by reducing the ratio of the coherences versus the fluctuations. 
By comparison with the straight black lines, which correspond to contour lines of $(N R)^2=[N \cot^2(\theta/2)]^2$, we indeed recover the quadratic behavior of the spin-coherence deviation except for smaller values of $N$ and close to $g^{(2)}(\bm{0}) = 2$, where the finite-$N$ deviation $-\frac{2}{N}$ (that is independent of the ratio between coherences and fluctuations) becomes relevant. 
\begin{figure}[t!]
\centering
\includegraphics[width=.5\textwidth]{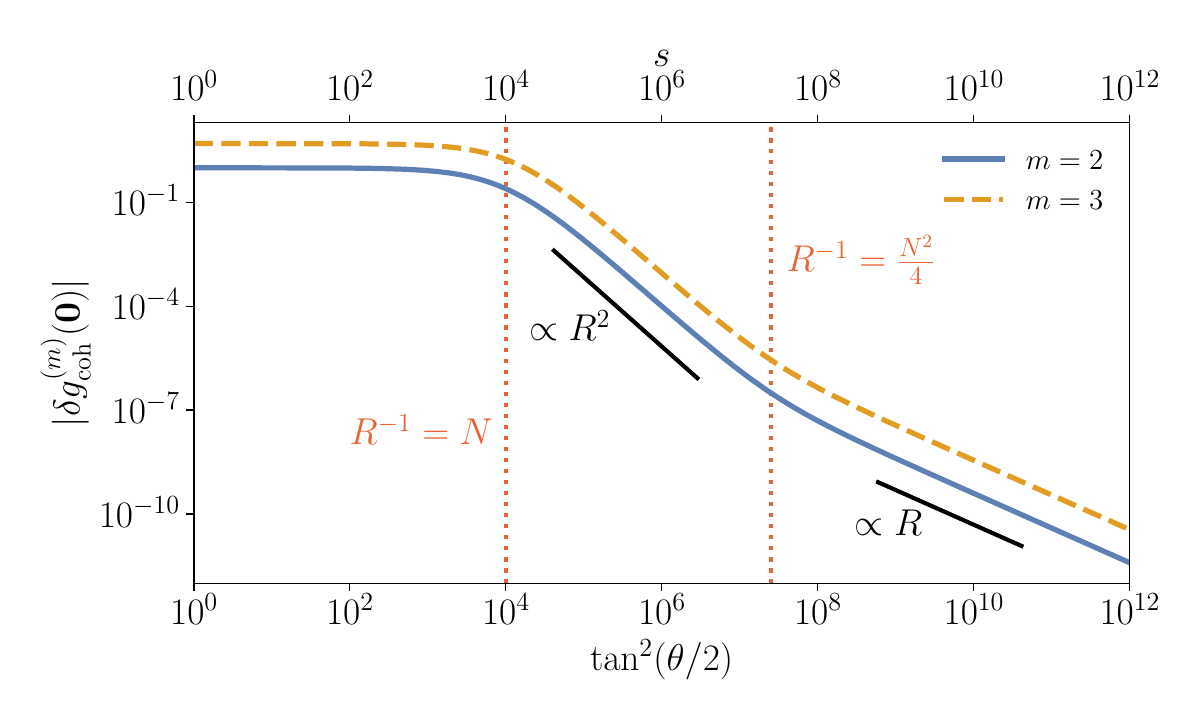}
\caption{Spin-coherence deviation of the $m$th-order autocorrelation function from the Gaussian Moment Theorem against the inverse ratio $R^{-1}=\tan^2(\theta/2)$ (bottom axis) and $R^{-1} = s$ (top axis), for $m \in \{2,3\}$ and $N=10^4$. Decreasing the ratio $R=\cot^2(\theta/2)= 1/s$ to $\approx \frac{1}{N}$, when spontaneous emission overtakes coherent emission, leads first to a regime in which the deviation scales quadratically in $R$ until the ratio reaches a value of $\approx \frac{4}{N^2}$. At this point, the first-order correction becomes dominant, so that the deviation $|\delta g^{(m)}_{\mathrm{coh}}(\bm{0})|$ scales linearly in $R$ afterwards (see App.~\ref{sec:AppA}).}
\label{Fig:gmm_diff_coh_sat}
\end{figure}

To analyze more precisely the dependence of the spin-coherence deviation on the ratio $R=\cot^2(\theta/2)$, we plot in Fig.~\ref{Fig:gmm_diff_coh_sat} the deviation for $m\in \{2,3\}$ and $N=10^4$ atoms against the inverse ratio $R^{-1} = \tan^2(\theta/2)$ (bottom axis). At first, the spin-coherence deviation stays constant until $R^{-1}=\tan^2(\theta/2)\approx N$, where spontaneous emission overtakes coherent emission. Then, the deviation drops quadratically according to the second-order correction until the ratio of the coherences and fluctuations reaches a value of $\approx 4/N^2$, when the first-order correction becomes dominant and the deviation scales linearly in $R$ afterwards (see App.~\ref{sec:AppA}).

\subsubsection{Continuously laser-driven atomic ensemble}\label{sec:Continuously_driven}

As a second example, we consider an atomic ensemble that is continuously driven by a plane wave laser with wave vector $\bm{k}_L$ and a saturation parameter $s$. The single-atom steady state is then given by 
\begin{align}
	\hat{\rho}_{\mu} = \begin{pmatrix}
		\frac{s}{2(1+s)} & -\frac{\sqrt{s}}{\sqrt{2}(1+s)}\\
		-\frac{\sqrt{s}}{\sqrt{2}(1+s)} & \frac{2+s}{2(1+s)}
	\end{pmatrix},
\label{eq:class_non_class_driven_atom_state}
\end{align}
for all $\mu \in \{1,...,N\}$, where $s$ denotes the saturation parameter. The atomic expectation values then read:
\begin{align}
	\label{eq:class_non_class_ssexp1}
	\braket{\hat{\sigma}^{+}\hat{\sigma}^{-}} &= \frac{s}{2(1+s)}\,,\\
	\braket{\hat{\sigma}^{-}} = \braket{\hat{\sigma}^{+}} &= -\frac{\sqrt{s}}{\sqrt{2}(1+s)}\,,\\
    \braket{\delta\hat{\sigma}^{+}\delta\hat{\sigma}^{-}} &= \frac{s^2}{2(1+s)^2}\,.
	\label{eq:class_non_class_ssexp4}
\end{align}
Correspondingly, the ratio of the coherences and fluctuations of the atoms is given by
\begin{align}
	\label{eq:class_non_class_coh_ratio}
	R = \frac{\braket{\hat{\sigma}^{+}} \braket{\hat{\sigma}^{-}}}{\braket{\delta\hat{\sigma}^{+}\delta\hat{\sigma}^{-}}} = \frac{1}{s},
\end{align}
being the inverse of the saturation parameter $s$.

Considering again the direction of observation $\bm{k}=\bm{0}$, the same dependencies as in the pulse-excited case are found for the spin-coherence deviation to the GMT, but now as a function of the saturation parameter $s$ (see Fig.~\ref{Fig:gmm_diff_coh_sat} top axis). 

\subsubsection{Off-axis scaling}
\label{sec:Off_axis_scaling}

\begin{figure}[t!]
\centering
\includegraphics[width=.5\textwidth]{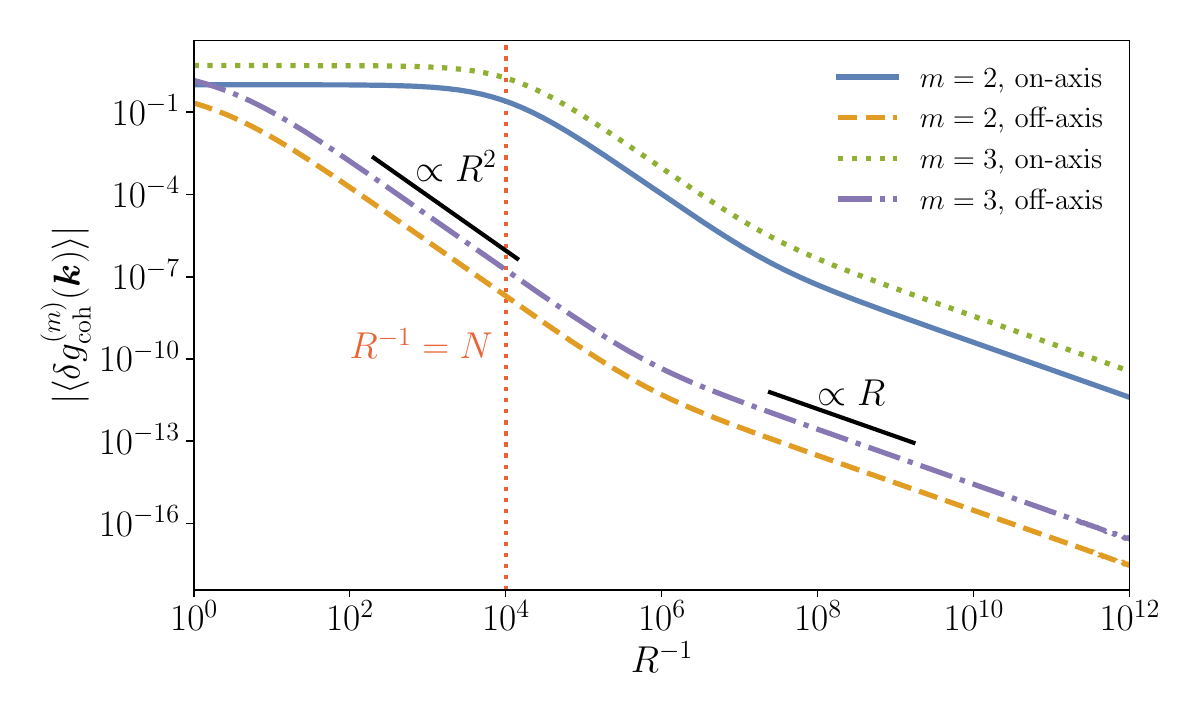}
\caption{Spin-coherence deviation $|\braket{\delta g^{(m)}_{\mathrm{coh}}(\bm{k})}|$ for $m\in\{2,3\}$, averaged over $1000$ realizations of the atomic ensemble, against the inverse ratio $R^{-1}$ for $N=10^4$ atoms. In contrast to the on-axis (laser direction) deviation, the off-axis deviation immediately decreases as $R^2$ becomes small compared to $1$. Thereby, the deviation scales quadratically in $R$ until the first-order correction overtakes, which is not exactly $0$ due to the finite number of realizations.}
\label{Fig:g22_g33_diff_sat_avg}
\end{figure}

Next, we still consider the autocorrelation function, but not along the direction $\bm{k}=\bm{0}$, i.e., in the direction of the laser, but in a random off-axis observation direction $\bm{k}_{\mathrm{obs}}\perp \bm{k}_L$. In this case, the intensity scales on average as $N$ instead of $N^2$. For example, this is the scaling observed when a randomization mechanism makes the intensity fluctuate with time. This does not change the condition for the finite-$N$ deviation $\delta g^{(m)}_{N}(\bm{k})$, but it loosens the $N$ dependence of the spin-coherence condition, which we discuss in detail in the case of the second- and third-order autocorrelation functions in the following. Hitherto, we perform a Taylor expansion in $\varepsilon=m!\sqrt{R}$ of the spin-coherence deviation $\delta g^{(m)}_{\mathrm{coh}}(\bm{k})$, which for $m=2$ gives
\begin{widetext}
\begin{align}
    \delta g^{(2)}_{\mathrm{coh}}(\bm{k}) = \frac{|S(\bm{k})|^2-N}{N^2} \varepsilon^2 - \frac{N|S(2\bm{k})|^2-(N-10)|S(\bm{k})|^4-8N|S(\bm{k})|^2-N[S(2\bm{k})S(-\bm{k})^2+S(-2\bm{k})S(\bm{k})^2]}{16N^3}\varepsilon^4 + \mathcal{O}(\varepsilon^6)\,.
\end{align}
\end{widetext}
Considering the ensemble averages $\braket{|S(\bm{k})|^2}\approx N$, $\braket{|S(\bm{k})|^4}\approx 2N^2$, $\braket{|S(2\bm{k})|^2}\approx N$, and $\braket{S(2\bm{k})S(-\bm{k})^2}=\braket{[S(-2\bm{k})S(\bm{k})^2]^*}\approx N$ for sufficiently large $N$, the spin-coherence deviation scales in this case as 
\begin{align}
    |\braket{\delta g^{(2)}_\mathrm{coh}(\bm{k})}| \approx \frac{2N-11}{16N}\varepsilon^4 + \mathcal{O}(\varepsilon^6) \approx 2 R^2 + \mathcal{O}(\varepsilon^6)\,.
\end{align}
A similar analysis for $m=3$ leads to
\begin{align}
    |\braket{\delta g^{(3)}_\mathrm{coh}(\bm{k})}| &\approx \frac{2N^2-19N+32}{144N^2}\varepsilon^4 + \mathcal{O}(\varepsilon^6)\nonumber\\
    &\approx 18 R^2 + \mathcal{O}(\varepsilon^6)\,.
\end{align}
Hence, the spin-coherence deviation becomes negligible in the limit where $R \to 0$, without any dependence on the number of atoms $N$. In Fig.~\ref{Fig:g22_g33_diff_sat_avg}, we plot the spin-coherence deviation, averaged over $1000$ different realizations of the atomic ensemble, against the inverse ratio $R^{-1}$. 
As one can see, the off-axis spin-coherence deviation indeed decreases as $R^2 \ll 1$ and scales quadratically in $R$. Further decreasing $R$ leads to a regime where the linear correction overtakes the quadratic one, since for a finite number of realizations the linear order correction is not exactly null.

\section{Conditions for the GMT in two-level systems for $m\neq n$\label{sec:Conditions_for_GMT_m_unequal_n}}

\subsection{Condition for the GMT}
\label{sec:Cond_for_GMT_m_unequal_n}

Since for $m \neq n$ there is always at least one field component whose phase is not canceled by the corresponding complex conjugate expression, all deviations from the GMT are connected to nonzero coherences. In other words, in the case $m \neq n$, there is no finite-$N$ condition, but only a spin-coherence condition reading
\begin{align}	\label{eq:spin_coh_cond_2}
    \sqrt{R} &\ll \frac{(N-x)! N^x}{x!N! \sqrt{N}} \approx \frac{1}{x! \sqrt{N}}\,,
\end{align}
where $x=\mathrm{max}\{m,n\}$. We note that due to the square root dependence, it is the most restrictive condition on the ratio of coherent to incoherent radiation $R$. However, we mention that, on the one hand, the $N$ dependence may again be loosened for an off-axis direction, and, on the other hand, the square root scaling in Eq.~\eqref{eq:spin_coh_cond_2} is the most restrictive scaling, valid for $|m-n|=1$, and is loosened for $|m-n|>1$ (see App.~\ref{sec:AppA} and examples below).

\subsection{Deviations from the GMT}
\label{sec:Dev_from_GMT_m_unequal_n}
 
\begin{figure}[t!]
\centering
\includegraphics[width=.5\textwidth]{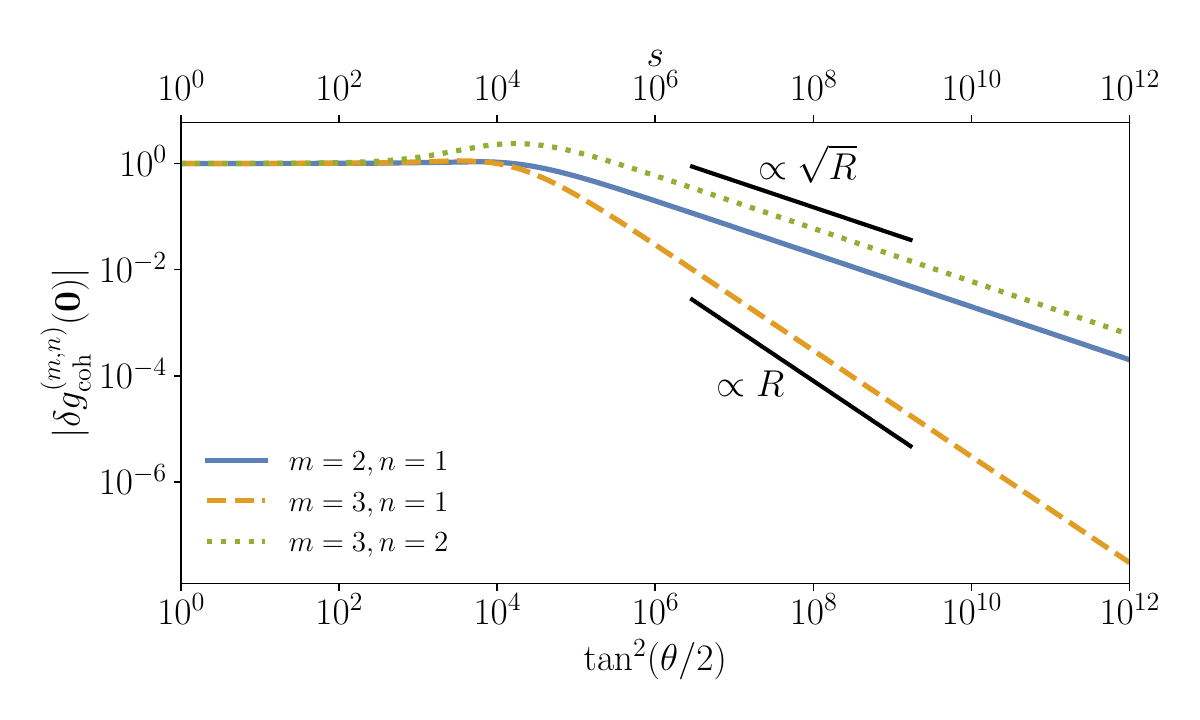}
\caption{Spin-coherence deviation $|\delta g^{(m,n)}_{\mathrm{coh}}(\bm{0})|$ against the inverse ratio $R^{-1}$ with $R^{-1}=\tan^2(\theta/2)$ (bottom axis) and $R^{-1} = s$ (top axis), for $N=10^4$ and $m \in \{2,3\}$, $n \in \{1,2\}$ ($m>n$). By decreasing the ratio $R$, the correlation functions approach the value $0$ given by the Gaussian Moment Theorem with different scalings. For $n=m-1$, the correlation functions scale as $\sqrt{R}$, for $n=m-2$ as $R$ (see Eq.~\eqref{eq:condSpSmgen} in App.~\ref{sec:AppA}).}
\label{Fig:gmn_coh_sat}
\end{figure}

Since for $m \neq n$ there are no deviations due to finite size effects, we can write 
\begin{align}
\delta g^{(m,n)}(\bm{k}_1,...,\bm{k}_{m+n}) 
&= \delta g^{(m,n)}_{\mathrm{coh}}(\bm{k}_1,...,\bm{k}_{m+n})\,.
\end{align}
In what follows, we consider $m>n$ without loss of generality. Following the spin-coherence condition~\eqref{eq:spin_coh_cond_2}, we use again the expansion parameter $\varepsilon = m!\sqrt{R}$. Performing Taylor expansions in $\varepsilon$ leads to [see Eq.~\eqref{eq:condSpSmgen} in App.~\ref{sec:AppA}]
\begin{align}
    |\delta g^{(2,1)}_{\mathrm{coh}}(\bm{0})| &\approx 2 \sqrt{N R} + \mathcal{O}(\varepsilon^3)\,,\\
    |\delta g^{(3,1)}_{\mathrm{coh}}(\bm{0})| &\approx 3 N R + \mathcal{O}(\varepsilon^4)\,,\\
    |\delta g^{(3,2)}_{\mathrm{coh}}(\bm{0})| &\approx 6 \sqrt{N R} + \mathcal{O}(\varepsilon^3)\,.
\end{align}
We highlight these behaviors in Fig.~\ref{Fig:gmn_coh_sat}, where we plot the deviations against $R^{-1}=\tan^2(\theta/2)$ for pulse-excited atoms (bottom axis) and $R^{-1}= s$ for continuously driven atoms (top axis).


\section{Comparison of quantum mechanical and classical deviations}
\label{sec:Comparison}

Let us finally compare our results derived for two-level emitters to those obtained for classical dipoles. Indeed, classical emitters can emit more than one photon at a time, which means that they do not exhibit quantum features such as antibunching for light. We derive the classical emission by considering that each particle $\mu$ at position $\bm{R}_\mu$ emits a field $e^{i\bm{k}\cdot\bm{R}_\mu} (E_\mathrm{coh} + E_\mathrm{incoh} e^{i\phi_\mu})$, with the first term being a coherent component and the second term being characterized by a fluctuating phase $\phi_\mu$ (for independent emitters the different phases $\phi_\mu$ are uncorrelated). These terms are the equivalent of the elastic and spontaneously emitted fields of a two-level atom.

As shown in App.~\ref{sec:AppA} and~\ref{sec:AppB}, the leading finite-$N$ deviation for the intensity correlations of order $m$ for classical emitters is given by $\delta g_{N}^{(m)}=-m!m(m-1)/4N$, thus being a factor $2$ smaller than its quantum counterpart [i.e., $-m!m(m-1)/2N$].  Similarly, we have verified that higher-order deviations also differ only by numerical factors (depending on the correlation order $m$, see App.~\ref{sec:AppA} and~\ref{sec:AppB}).

Concerning the deviation due to spin-coherence, the leading term in the large $N$ limit is actually the same for two-level atoms and classical emitters (see the $\propto R^2$ and $\sqrt{R}$ scalings discussed in Figs.~\ref{Fig:gmm_diff_coh_sat} and~\ref{Fig:gmn_coh_sat}). However, for $m=n$, the deviation in $R$ differs again by a factor of $2$, in addition to a minus sign (see App.~\ref{sec:AppA} and~\ref{sec:AppB}). These results highlight that the two-level nature of emitters may actually be encapsulated in the deviations of the light statistics from the GMT.


\section{Conclusion}
\label{sec:Conclusion}

In this work, we have shown that for systems of uncorrelated emitters, deviations from the Gaussian momentum theorem occur either from finite-size effects or from the presence of spin coherence. Indeed, the latter creates an interference pattern, even without the emitters interacting, which prevents the validity of the theorem. We have thus derived two conditions which must be fulfilled to observe the thermal light statistics characterized by the Gaussian moment theorem for ensembles of two-level atoms in a product state, one restricting the order of the correlation function with respect to the number of atoms, and another restricting the ratio of the single-atom coherence with respect to its fluctuations. We have provided, at leading orders, the corrections to the values predicted by the Gaussian moment theorem for the second- and third-order correlation functions, considering as concrete examples fully excited states, coherent spin states and laser-driven steady states. Finally, we pointed out that the corrections differ quantitatively for quantum (two-level) emitters and classical dipoles, due to the inability of the former to emit two photons at a time; hence, the deviations from thermal statistics in ensembles of motionless emitters may be used to probe the quantized nature of the level structure of the atoms. Our work helps understanding which deviations are to be expected from thermal statistics, either from particle numbers~\cite{Watts1996} or from coherent components~\cite{Boas1997, Borycki2016}, without necessarily stemming from interactions~\cite{Voigt1994,Ferioli2024}.

\begin{acknowledgments}
R.B. and J.v.Z. gratefully acknowledge funding and support by the  Bavarian Academic Center for Latin America (BAYLAT). This work was funded by the Deutsche Forschungsgemeinschaft (DFG, German Research Foundation) -- Project-ID 429529648 -- TRR 306 QuCoLiMa (``Quantum Cooperativity of Light and Matter''). A.C. and R.B. have received the financial support of the S\~ao Paulo Research Foundation (FAPESP) (Grants No. 2022/00209-6, 2022/06449-9, 2023/07463-8, 2025/00697-9 and 2023/03300-7) and from the Brazilian CNPq (Conselho Nacional de Desenvolvimento Científico e Tecnológico), Grant No. 313632/2023-5 and 403788/2025-0.
\end{acknowledgments}

\bibliography{GMT.bib}

\appendix

\section{Quantum mechanical derivation of the conditions}
\label{sec:AppA}
\subsection{Case $m=n$}
We start with the case $m=n$. Since $(\hat{\sigma}^\pm_\mu)^l = 0$ for $l\geq 2$, we can write
\begin{align}
	&G^{(m)}(\bm{k}_1,...,\bm{k}_m,\bm{k}_{m+1},...,\bm{k}_{2m}) =\nonumber\\
    &\sum_{\substack{\mu_1,...,\mu_m=1,\\ \text{mut. diff.}}}^{N} \sum_{\substack{\nu_1,...,\nu_m=1,\\ \text{mut. diff.}}}^{N} e^{i \bm{k}_1. \bm{R}_{\mu_1}}...e^{i \bm{k}_m. \bm{R}_{\mu_m}}e^{-i \bm{k}_{m+1}. \bm{R}_{\nu_m}}...e^{-i \bm{k}_{2m}. \bm{R}_{\nu_1}}\times\nonumber\\
    &\times\braket{\hat{\sigma}^+_{\mu_1}...\hat{\sigma}^+_{\mu_m}\hat{\sigma}^-_{\nu_m}...\hat{\sigma}^-_{\nu_1}}\nonumber\\
	&= \sum_{\substack{\mu_1,...,\mu_m=1,\\ \text{mut. diff.}}}^{N} \sum_{\nu_1<...<\nu_m=1}^{N}\sum_{\sigma \in S_m} \prod_{\{p,q\}\in P_{\sigma}} e^{i \bm{k}_p. \bm{R}_{\mu_p}} e^{-i \bm{k}_{q}. \bm{R}_{\nu_p}} \braket{\hat{\sigma}^+_{\mu_p}\hat{\sigma}^-_{\nu_p}}\,,
	\label{eq:class_non_class_Gm}
\end{align}
where the abbreviation mut. diff. stands for mutually different. On the other hand, the first-order correlation function reads
\begin{align}
	G^{(1)}(\bm{k}_1,\bm{k}_2) = \sum_{\mu,\nu=1}^{N} e^{i \bm{k}_1. \bm{R}_{\mu}} e^{-i \bm{k}_2. \bm{R}_{\nu}} \braket{\hat{\sigma}^+_{\mu} \hat{\sigma}^-_{\nu}}\,,
\end{align}
so that
\begin{align}
	\label{eq:class_non_class_Gausstot}
	&\sum_{\sigma \in S_m} \prod_{\{p,q\}\in P_{\sigma}} G^{(1)}(\bm{k}_p,\bm{k}_q)\nonumber\\ 
    &= \sum_{\sigma \in S_m} \prod_{\{p,q\}\in P_{\sigma}} \sum_{\mu,\nu=1}^{N} e^{i \bm{k}_p. \bm{R}_{\mu}} e^{-i \bm{k}_{q}. \bm{R}_{\nu}} \braket{\hat{\sigma}^+_{\mu} \hat{\sigma}^-_{\nu}}\nonumber\\
    &= \sum_{\mu_1,...,\mu_m=1}^{N} \sum_{\nu_1,...,\nu_m=1}^{N} \sum_{\sigma \in S_m} \prod_{\{p,q\}\in P_{\sigma}} e^{i \bm{k}_p. \bm{R}_{\mu_p}} e^{-i \bm{k}_{q}. \bm{R}_{\nu_p}} \braket{\hat{\sigma}^+_{\mu_p} \hat{\sigma}^-_{\nu_p}}\,.
\end{align}
Comparing Eq.~\eqref{eq:class_non_class_Gm} to Eq.~\eqref{eq:class_non_class_Gausstot}, we find two differences, namely the different sums
\begin{align}
	\sum_{\substack{\mu_1,...,\mu_m=1,\\ \text{mutually different}}}^{N} &\leftrightarrow \sum_{\mu_1,...,\mu_m=1}^{N}\,,\\
	\sum_{\nu_1<...<\nu_m=1}^{N} &\leftrightarrow \sum_{\nu_1,...,\nu_m=1}^{N}\,.
\end{align}
To check whether we can potentially obtain the Gaussian Moment Theorem, we define the sums 
\begin{align}
	\Sigma_1({N},m) &\coloneqq \sum_{\mu_1,...,\mu_m=1}^{N} 1 = {N}^m\,,\\
	\Sigma_2({N},m) &\coloneqq \sum_{\substack{\mu_1,...,\mu_m=1,\\ \text{mutually different}}}^{N} 1 = m! \binom{{N}}{m}\,,\\
	\Sigma_3({N},m) &\coloneqq \sum_{\mu_1<...<\mu_m=1}^{N} 1 = \binom{{N}}{m}\,,
\end{align}
and calculate the relative errors of the first and second and of the first and third sum
\begin{align}
	&\lim_{{N}\rightarrow \infty} \left(\frac{\Sigma_1({N},m)-\Sigma_2({N},m)}{\Sigma_1({N},m)}\right) = 0\,,\\
	&\lim_{{N}\rightarrow \infty} \left(\frac{\Sigma_1({N},m)-\Sigma_3({N},m)}{\Sigma_1({N},m)}\right) = 1 - \frac{1}{m!}\,.
\end{align}
The relative error between the first two sums vanishes for $N \rightarrow \infty$, which suggests a proper approximation of the first sum by the second sum. However, the relative error between the first and the third sum is finite for $m>2$, which means that we need to be able to neglect certain summands. Therefore, we note that the difference in the correlation functions is only present if there are nonzero single-atom coherences $\braket{\hat{\sigma}^\pm_{\mu}}\neq 0$. If it would hold that $\braket{\hat{\sigma}^+_{\mu_p} \hat{\sigma}^-_{\nu_p}} = \braket{\hat{\sigma}^+_{\mu_p} \hat{\sigma}^-_{\mu_p}}\delta_{\mu_p,\nu_p} = \braket{\delta\hat{\sigma}^+_{\mu_p} \delta\hat{\sigma}^-_{\mu_p}}\delta_{\mu_p,\nu_p}$, the sums over the $\nu$s disappear and we can potentially obtain the Gaussian Moment Theorem. Here, $\delta\hat{\sigma}_\mu^\pm = \hat{\sigma}_\mu^\pm-\braket{\hat{\sigma}_\mu^\pm}$ describes the incoherent fluctuations. This will lead us to a condition relating the single-atom coherences with respect to the single-atom fluctuations. Let us therefore investigate the ratio of the coherences and the fluctuations of the atoms. Since the deviation results from the coherent contribution of the radiation, which constructively interferes and thus is largest for $\bm{k}=\bm{0}$, we consider the direction $\bm{k}=\bm{0}$ in what follows. Here, we have
\begin{align}
	&\!\!\!G^{(m)}(\bm{0},...,\bm{0}) = \sum_{\mu_1,...,\mu_m,\nu_1,...,\nu_m=1}^{N} \braket{\hat{\sigma}^+_{\mu_1}...\hat{\sigma}^+_{\mu_m}\hat{\sigma}^-_{\nu_m}...\hat{\sigma}^-_{\nu_1}}\nonumber\\
	=& \sum_{j=0}^{m} \binom{m}{j}^2 j! (2m-j)! \binom{{N}}{2m-j} (\braket{\hat{\sigma}^+ \hat{\sigma}^-})^j \left(\braket{\hat{\sigma}^+}\braket{\hat{\sigma}^-}\right)^{m-j}\nonumber\\
    =& \sum_{j=0}^{m} \binom{m}{j}^2 j! (2m-j)! \binom{{N}}{2m-j}\nonumber\\
    &\times\sum_{l=0}^{j} \binom{j}{l}  (\braket{\delta\hat{\sigma}^+ \delta\hat{\sigma}^-})^l \left(\braket{\hat{\sigma}^+}\braket{\hat{\sigma}^-}\right)^{m-l}\,,
	\label{eq:class_non_class_Forward}
\end{align}
where we neglected the atomic index since all atoms are assumed to be in the same state. To determine under which condition the contributions from the coherences can be neglected compared to those from the fluctuations, let us calculate the ratio between the terms corresponding to $l=m-q$ and $l=m-q+1$, which is given by
\begin{align}
	\frac{\braket{\hat{\sigma}^+}\braket{\hat{\sigma}^-}}{\braket{\delta\hat{\sigma}^+ \delta\hat{\sigma}^-}}\frac{(N-m+q)(m-q+1)}{q^2}\,.
\end{align}
Since we want to keep only the term with $q=0$, we demand this ratio to be much smaller than $1$ for $q=1$. Therefore, we require
\begin{align}
	&\frac{\braket{\hat{\sigma}^+}\braket{\hat{\sigma}^-}}{\braket{\delta\hat{\sigma}^+ \delta\hat{\sigma}^-}} m(N-m+1) \ll 1\nonumber\\
    &\Leftrightarrow \frac{\braket{\hat{\sigma}^+}\braket{\hat{\sigma}^-}}{\braket{\delta\hat{\sigma}^+ \delta\hat{\sigma}^-}} \ll \frac{1}{m(N-m+1)}\,,
	\label{eq:class_non_class_s_req}
\end{align}
which also implies that all other terms with $q>1$ can be neglected. However, we note that actually a weaker condition is already sufficient if one considers the normalized correlation function. This can be seen by performing a Taylor expansion of the normalized correlation function $g^{(m)}(\bm{0},...,\bm{0})$ in orders of $R = \frac{\braket{\hat{\sigma}^+}\braket{\hat{\sigma}^-}}{\braket{\delta\hat{\sigma}^+ \delta\hat{\sigma}^-}}$ given by
\begin{align}
	\label{eq:App_GMT_gm_Taylor}
	&g^{(m)}(\bm{0},...,\bm{0}) = \frac{(m!)^2 \binom{N}{m}}{N^m} - \frac{(m!)^2 \binom{N}{m}m(m-1)}{N^{m}}R\nonumber\\
	&- \frac{1}{4} \frac{(m!)^2 \binom{N}{m}(N^2-3N-2mN+3m-m^2-2)m(m-1)}{N^{m}}R^2\nonumber\\
    &+ \mcal{O}(R^3)\,. 
\end{align}
Later, we will require that $\frac{m!m (m-1)}{2N} \ll 1$, so that $\frac{m! \binom{N}{m}}{N^m}\approx 1$ and the correlation function can be approximated by
\begin{align}
	\label{eq:App_GMT_gm_Taylor_large_N}
	g^{(m)}(\bm{0},...,\bm{0}) \approx\;& m! - m! m (m-1)R\nonumber\\
	&- \frac{1}{4} m! m (m-1)N^2 R^2 + \mcal{O}(R^3)\,. 
\end{align}
Therefore, the second-order correction will be dominant until $R \approx \frac{4}{N^2}$, for which the first-order correction takes over. This also means that we only need the weaker condition
\begin{align}
\label{eq:condsq}
    \left(\frac{\braket{\hat{\sigma}^+}\braket{\hat{\sigma}^-}}{\braket{\delta\hat{\sigma}^+ \delta\hat{\sigma}^-}}\right)^2 \ll \frac{4}{N^2 m (m-1)}\,. 
\end{align}
However, we also note that this is only valid for $m=n$. As we will find out later, the case $m\neq n$ demands a condition for $\sqrt{R}$, stronger than the linear~\eqref{eq:class_non_class_s_req} and quadratic~\eqref{eq:condsq} ones.\\
Nevertheless, if the condition~\eqref{eq:class_non_class_s_req} is fulfilled, we can approximate Eq.~\eqref{eq:class_non_class_Gm} in leading order by
\begin{align}
	\label{eq:class_non_class_Gm_approx}
	&G^{(m)}(\bm{k}_1,...,\bm{k}_m,\bm{k}_{m+1},...,\bm{k}_{2m})\nonumber\\
 &\approx \sum_{\substack{\mu_1,...,\mu_m=1,\\ \text{mutually different}}}^{N} \sum_{\sigma \in S_m} \prod_{\{p,q\}\in P_{\sigma}} e^{i(\bm{k}_p-\bm{k}_{q}).\bm{R}_{\mu_p}} \braket{\delta\hat{\sigma}^+_{\mu_p} \delta\hat{\sigma}^-_{\mu_p}}\nonumber\\
 &=\braket{\delta\hat{\sigma}^+ \delta\hat{\sigma}^-}^m \sum_{\substack{\mu_1,...,\mu_m=1,\\ \text{mutually different}}}^{N} \sum_{\sigma \in S_m} \prod_{\{p,q\}\in P_{\sigma}} e^{i(\bm{k}_p-\bm{k}_{q}).\bm{R}_{\mu_p}}
\end{align}
and Eq.~\eqref{eq:class_non_class_Gausstot} by
\begin{align}
	\label{eq:class_non_class_Gauss_approx}
	&\sum_{\sigma \in S_m} \prod_{\{p,q\}\in P_{\sigma}} G^{(1)}(\bm{k}_p,\bm{k}_{q})\nonumber\\ &\approx \sum_{\mu_1,...,\mu_m=1}^{N} \sum_{\sigma \in S_m} \prod_{\{p,q\}\in P_{\sigma}} e^{i(\bm{k}_p-\bm{k}_{q}).\bm{R}_{\mu_p}} \braket{\delta\hat{\sigma}^+_{\mu_p} \delta\hat{\sigma}^-_{\mu_p}}\nonumber\\
    &=\braket{\delta\hat{\sigma}^+ \delta\hat{\sigma}^-}^m \sum_{\mu_1,...,\mu_m=1}^{N} \sum_{\sigma \in S_m} \prod_{\{p,q\}\in P_{\sigma}} e^{i(\bm{k}_p-\bm{k}_{q}).\bm{R}_{\mu_p}}\,.
\end{align}
In particular, Eq.~\eqref{eq:class_non_class_Gm_approx} means that we can approximate
\begin{align}
	G^{(1)}(\bm{k},\bm{k}) \approx N \braket{\delta\hat{\sigma}^+ \delta\hat{\sigma}^-}\,,
\end{align}
which leads to 
\begin{align}
	\label{eq:class_non_class_gm_approx}
	&g^{(m)}(\bm{k}_1,...,\bm{k}_m,\bm{k}_{m+1},...,\bm{k}_{2m})\approx\nonumber\\
 & \frac{1}{N^m} \sum_{\substack{\mu_1,...,\mu_m=1,\\ \text{mutually different}}}^{N} \sum_{\sigma \in S_m} \prod_{\{p,q\}\in P_{\sigma}} e^{i(\bm{k}_p-\bm{k}_{q}).\bm{R}_{\mu_p}}\,.
\end{align}
To account for the mutual difference of the indices of summation $\mu_1,...,\mu_{m}$ in Eq.~\eqref{eq:class_non_class_gm_approx}, we include the following product
\begin{align}
	&\prod_{\nu_1 \in \{ \mu_{2},...,\mu_{m}\}}\!\!(1-\delta_{\mu_1,\nu_1}) \times \!\!\!\prod_{\nu_2 \in \{ \mu_{3},...,\mu_{m}\}}\!\!(1-\delta_{\mu_2,\nu_2}) \times\! ... \!\times (1-\delta_{\mu_{m-1},\mu_m})\nonumber\\
	&= 1 + f(\delta_{\mu_1,\mu_2},...,\delta_{\mu_1,\mu_m},\delta_{\mu_2,\mu_3},...,\delta_{\mu_2,\mu_m},...,\delta_{\mu_{m-1},\mu_m})\,,
 \label{eq:delta_func}
\end{align}
where $f(\delta_{\mu_1,\mu_2},...,\delta_{\mu_1,\mu_m},\delta_{\mu_2,\mu_3},...,\delta_{\mu_2,\mu_m},...,\delta_{\mu_{m-1},\mu_m})$ is a multivariate polynomial of degree $m-1$. We note that the lowest monomial of $f$ has a degree of 1.\\
Since the first summand in Eq.~\eqref{eq:delta_func} gives the Gaussian Moment Theorem, all contributions coming from the $f$ function shall be small compared to the contribution from the first summand. However, we do not demand the contributions from the $f$ function to be relatively small, but absolutely small. Therefore, we use the upper estimate
\begin{align}
	&\left|\sum_{\mu_1,...,\mu_{m}=1}^{N} \delta_{\mu_s,\mu_t} \sum_{\sigma \in S_m} \prod_{\{p,q\} \in P_{\sigma}} e^{i (\bm{k}_p-\bm{k}_q). \bm{R}_{\mu_p}} \right|\nonumber\\
    &\leq \sum_{\mu_1,...,\mu_{m}=1}^{N} \delta_{\mu_s,\mu_t} \sum_{\sigma \in S_m} \prod_{\{p,q\} \in P_{\sigma}} \left| e^{i (\bm{k}_p-\bm{k}_q). \bm{R}_{\mu_p}} \right| = m! N^{m-1}\,.
\end{align}
Therefore, the terms coming from the function $f$ are at most of order $N^{m-1}$. In addition, we also need to take into account the number of terms of a specific order of $N$. Therefore, let us calculate
\begin{align}
	\label{eq:class_non_class_counting}
	\sum_{\mu_1,...,\mu_m=1}^{N} (1+f) = m!\binom{N}{m}= N (N-1)...(N-m+1)\,,
\end{align}
which gives a falling factorial, a polynomial in $N$. The coefficients of this polynomial are the so-called Stirling numbers of the first kind. Thus, the number of terms of order $N^{l}$ is exactly given by the coefficient, i.e., the corresponding Stirling number of the $N^l$ term for $l \in \{1,...,m-1\}$ in the falling factorial. Let $\mcal{L}=\{-1,...,-m+1\}$, then the Stirling number of the first kind for a given $m$ and $l$ is~\cite{Stanley_2011}
\begin{align}
	\label{eq:class_non_class_coeff}
	\mathcal{S}(m,l) = \sum_{\substack{T \subset \mcal{L},\\ |T|=m-l}} \prod_{p=1}^{m-l} T_p = (-1)^{m-l} \sum_{0 \leq i_1<i_2<...<i_{m-l}\leq m-1} \prod_{p=1}^{m-l} i_p\,.
\end{align}
Since
\begin{align}
    |\mathcal{S}(m,m-1)|^l &= \sum_{i_l=0}^{m-1}...\sum_{i_1=0}^{m-1} i_l...i_1 > \sum_{i_l=0}^{m-1}\sum_{i_{l-1}=0}^{i_l-1}...\sum_{i_1=0}^{i_2-1} i_l...i_1\nonumber\\
    &= |\mathcal{S}(m,m-l)|\,,
\end{align}
we obtain for the corrections
\begin{align}
&\left|\frac{1}{N^m} \sum_{\mu_1,...,\mu_m=1}^{N} f \sum_{\sigma \in S_m} \prod_{\{p,q\}\in P_{\sigma}} e^{i(\bm{k}_p-\bm{k}_{q}).\bm{R}_{\mu_p}} \right|\nonumber\\
&< m!\left(\frac{|\mathcal{S}(m,m-1)|}{N} + \frac{|\mathcal{S}(m,m-1)|^2}{N^2} + ... + \frac{|\mathcal{S}(m,m-1)|^{m-1}}{N^{m-1}}\right)
\label{eq:class_non_class_Order}
\end{align}
Therefore, it is justified to keep only the zeroth order of the $m$th-order correlation function if $\frac{m! |\mathcal{S}(m,m-1)|}{N} = \frac{m! m (m-1)}{2N} \ll 1$. However, we note that it is usually enough to have $\frac{m (m-1)}{2N} \ll 1$ since also the terms contributing to the Gaussian Moment Theorem are $m!$ many. Finally, using Eq.~\eqref{eq:class_non_class_gm_approx} and approximating the normalized $m$th-order correlation function in zeroth order gives
\begin{widetext}
\begin{align}
g^{(m)}(\bm{k}_1,...,\bm{k}_{2m}) &\approx \frac{\braket{\delta\hat{\sigma}^+ \delta\hat{\sigma}^-}^m}{\left(N \braket{\delta\hat{\sigma}^+ \delta\hat{\sigma}^-}\right)^m} \sum_{\mu_1,...,\mu_m=1}^{N} \sum_{\sigma \in S_m} \prod_{\{p,q\}\in P_{\sigma}} e^{i(\bm{k}_p-\bm{k}_{q}).\bm{R}_{\mu_p}} \nonumber \\
	&\approx \sum_{\sigma \in S_m} \prod_{\{p,q\}\in P_{\sigma}} \frac{G^{(1)}(\bm{k}_p,\bm{k}_q)}{\sqrt{G^{(1)}(\bm{k}_p,\bm{k}_p)}\sqrt{G^{(1)}(\bm{k}_q,\bm{k}_q)}} = \sum_{\sigma \in S_m} \prod_{\{p,q\}\in P_{\sigma}} g^{(1)}(\bm{k}_p,\bm{k}_{q})\,,
\end{align}
\end{widetext}
which is the Gaussian Moment Theorem.
\subsection{Case $m\neq n$}
Let us now discuss the case $m\neq n$. Considering the Gaussian Moment Theorem Eq.~\eqref{eq:GMT_Corr}, for $m\neq n$ we require the generalized higher-order correlation functions to be $0$. If the single atom states $\hat{\rho}_{\mu}$ do not possess any coherences, i.e., $\braket{\hat{\sigma}^\pm_{\mu}} = 0$, then $g^{(m,n)} = 0$ and the Gaussian Moment Theorem is fulfilled. However, if $\braket{\hat{\sigma}^\pm_{\mu}} \neq 0$, we require the coherences $\braket{\hat{\sigma}^\pm_{\mu}}$ to be sufficiently small, which we specify in the following. Since the deviation from the Gaussian Moment Theorem results from the coherent contribution of the radiation, which is largest for $\bm{k}=\bm{0}$, we consider the direction $\bm{k}=\bm{0}$ in what follows. The generalized higher-order correlation functions thus reduce to
\begin{widetext}
\begin{align}
	G^{(m,n)}(\bm{0},...,\bm{0}) = \sum_{j=0}^{\alpha} \binom{m}{j}\binom{n}{j} j! (2\alpha-j)! \binom{N}{2\alpha-j} (m-\alpha)! \binom{N-(2\alpha-j)}{m-\alpha} (n-\alpha)! \binom{N-(2\alpha-j)}{n-\alpha} \sum_{l=0}^{j} \binom{j}{l} \braket{\delta\hat{\sigma}^+ \delta\hat{\sigma}^-}^{l} \braket{\hat{\sigma}^+}^{m-l} \braket{\hat{\sigma}^-}^{n-l}\,,
\end{align}
where $\alpha \coloneqq \min\{m,n\}$. Let us wlog assume that $m>n$, then $\alpha=n$ and we can simplify
\begin{align}
	G^{(m,n)}(\bm{0},...,\bm{0}) = \sum_{j=0}^{n} \binom{m}{j}\binom{n}{j} j! (2n-j)! \binom{N}{2n-j} (m-n)! \binom{N-(2n-j)}{m-n} \sum_{l=0}^{j} \binom{j}{l} \braket{\delta\hat{\sigma}^+ \delta\hat{\sigma}^-}^{l} \braket{\hat{\sigma}^+}^{m-l} \braket{\hat{\sigma}^-}^{n-l}\,.
\end{align}
\end{widetext}
Note that if $m<n$, we would simply replace $m\leftrightarrow n$ and $\braket{\hat{\sigma}^+}\leftrightarrow \braket{\hat{\sigma}^-}$. To find the conditions for the coherences, we calculate the ratio of the terms corresponding to $l=n-q$ and $l=n-q+1$, which is given by
\begin{align}
    \frac{\braket{\hat{\sigma}^+} \braket{\hat{\sigma}^-}}{\braket{\delta\hat{\sigma}^+ \delta\hat{\sigma}^-}} \frac{(N-n+q)(n-q+1)}{q(m-n+q)}\,.
\end{align}
This ratio shall be much smaller than $1$ for all $q$. The strongest constraint is thus given for $q=1$ and $n=m-1$, so that we find the condition
\begin{align}
    \frac{\braket{\hat{\sigma}^+} \braket{\hat{\sigma}^-}}{\braket{\delta\hat{\sigma}^+ \delta\hat{\sigma}^-}} \ll \frac{2}{(m-1)(N-m+2)}\,.
\end{align}
Assuming this condition to be fulfilled, we approximate the normalized higher-order correlation function by the zeroth order as
\begin{align}
    g^{(m,n)}(\bm{0},...,\bm{0}) \approx \frac{\braket{\hat{\sigma}^+}^{m-n}}{\braket{\delta\hat{\sigma}^+ \delta\hat{\sigma}^-}^{\frac{m-n}{2}}} \frac{m!N!}{(m-n)!(N-m)!N^{\frac{m+n}{2}}}
\end{align}
Therefore, we require
\begin{align}
	&|g^{(m,n)}(\bm{0},...,\bm{0})| \approx \left(\frac{\braket{\hat{\sigma}^+} \braket{\hat{\sigma}^-}}{\braket{\delta\hat{\sigma}^+ \delta\hat{\sigma}^-}}\right)^{\frac{m-n}{2}} \frac{m!N!}{(m-n)!(N-m)!N^{\frac{m+n}{2}}} \ll 1\nonumber\\
 &\Leftrightarrow \left(\frac{\braket{\hat{\sigma}^+} \braket{\hat{\sigma}^-}}{\braket{\delta\hat{\sigma}^+ \delta\hat{\sigma}^-}}\right)^{\frac{m-n}{2}} \ll \frac{(m-n)!(N-m)! N^{\frac{m+n}{2}}}{m! N!}\,.
 \label{eq:condSpSmgen}
\end{align}
The strongest condition is again for $n=m-1$, for which we can simplify the condition as
\begin{align}
    \sqrt{\frac{\braket{\hat{\sigma}^+} \braket{\hat{\sigma}^-}}{\braket{\delta\hat{\sigma}^+ \delta\hat{\sigma}^-}}} \ll \frac{(N-m)! N^m}{m!N! \sqrt{N}} \approx \frac{1}{m! \sqrt{N}}\,,
    \label{eq:condSpSm}
\end{align}
which gives the most restrictive condition on the ratio of the coherences and the fluctuations.

\section{Classical derivation of the conditions}
\label{sec:AppB}
\title{Scaling of Higher-Order Autocorrelations for Classical Oscillators and Two-Level Atoms}
\subsection{Case $m=n$}
The total electric field emitted by $N$ classical oscillators at positions $\bm{R}_1,...,\bm{R}_N$ is composed of a  coherent (static) field component and a fluctuating incoherent field component, so that it can be written as
\begin{align}
	E(\bm{k}) = \sum_{\mu=1}^{N} e^{i\bm{k}.\bm{R}_\mu} \left(E_\mathrm{coh} + E_\mathrm{incoh} e^{i\phi_\mu}\right)\,.
\end{align}
Thereby, $E_\mathrm{coh}$ and $E_\mathrm{incoh}$ are constant and the same for each oscillator and $\phi_\mu$ is an oscillator-specific fluctuating phase. Since the coherent component is largest for $\bm{k}=\bm{0}$, we consider this direction in the following.\\
In the classical case the fields commute, so that the $m$th-order intensity correlation function can be written as
\begin{align}
    g^{(m)}(\bm{0}) = \frac{\braket{[I(\bm{0})]^m}}{\braket{I(\bm{0})}^m}\,.
\end{align}
The intensity can easily be calculated to be
\begin{align}
    \braket{I(\bm{0})} = N\left(|E_\mathrm{incoh}|^2 + N |E_\mathrm{coh}|^2 \right)\,,
\end{align}
whereas the expectation value $\braket{[I(\bm{0})]^m}$ gives
\begin{align}
    \braket{[I(\bm{0})]^m} = \sum_{j=0}^{m} \binom{m}{j}^2 (N^2 |E_\mathrm{coh}|^2)^{m-j}(|E_\mathrm{incoh}|^2)^j C_{j,N}\,.
\end{align}
Here,
\begin{align}
    C_{j,N} = \sum_{\mu_1,\dots,\mu_j,\,\nu_1,\dots,\nu_j = 1}^N \left\langle e^{i\sum_{p=1}^j (\phi_{\mu_p} - \phi_{\nu_p})} \right\rangle
\end{align}
involves expectations of exponentials of sums of phases which represent $j$ emission and $j$ detection events, corresponding to indices $\mu_p$ and $\nu_p$, respectively. The important point is that only terms where the total phase cancels will contribute to the ensemble average, i.e.,
\begin{equation}\label{Eq:ClassicalCondition}
\left\langle e^{i\sum_{p=1}^j (\phi_{\mu_p} - \phi_{\nu_p})} \right\rangle \ne 0 \iff \{\mu_1, \dots, \mu_j\} = \{\nu_1, \dots, \nu_j\},
\end{equation}
with $\{\cdot\}$ standing for unordered multisets of indices, i.e., sets that allow repeated elements. 
Note that condition \eqref{Eq:ClassicalCondition} leads to a purely combinatorial counting problem, as long as phase cancellation is achieved. 

\noindent Let a partition $\lambda$ of $j$, denoted by
\begin{equation}
\lambda = (1^{r_1}, 2^{r_2}, \dots, j^{r_j}), \quad \text{such that } \sum_{n=1}^{j} n r_n = j,
\end{equation}
encode one possible index multiset configuration $\{\mu_1, \dots, \mu_j\}$. Now, the general strategy to calculate the number of terms that survive in $C_{j,N}$ is to:
\begin{enumerate}
  \item Enumerate all integer partitions $\lambda$ of $j$ (i.e., $\lambda \vdash j$);
  \item For each partition $\lambda$, count the number of distinct configurations $B_{\lambda,N}$ that realize the corresponding multiplicities. This number is given by
  \begin{equation}
  B_{\lambda,N} = \frac{N!}{[N - \ell(\lambda)]! \prod_n r_n!},
  \end{equation}
where $\ell(\lambda)=\sum_nr_n$ is the total number of nonzero elements of the partition;
  
\item For each partition $\lambda$, account for the possible index permutations $P_{\lambda}$ in the multiset of indices $\mu$, which follows
 \begin{equation}
  P_{\lambda} = \frac{j!}{\prod_n (n!)^{r_n}}.
  \end{equation}
\end{enumerate}
The total number of surviving terms is thus given by
\begin{equation}
C_{j,N} =\sum_{\lambda \vdash j} C_{j,N}^{(\lambda)}= \sum_{\lambda \vdash j} B_{\lambda,N}  P_{\lambda}^2,
\end{equation}
where the square takes into account permutations in both $\mu$ and $\nu$ multisets. 

Finally, the $m$th-order intensity correlation function for the light scattered by $N$ classical emitters is given by
\begin{equation}
    g^{(m)}(\bm{0}) = \frac{1}{N^m(1+NR)^m} \sum_{j=0}^{N} \binom{m}{j}^2 (N^2 R)^{m-j} C_{j,N}\,,
\end{equation}
where $R = \frac{|E_\mathrm{coh}|^2}{|E_\mathrm{incoh}|^2}$ is equivalently defined as in the quantum mechanical case. We again perform a Taylor expansion up to the second order in $R$, which gives
\begin{align}
    &g^{(m)}(\bm{0}) = \frac{C_{m,N}}{N^m} + \frac{m^2 N^2 C_{m-1,N}-m N C_{m,N}}{N^m}R+\frac{1}{4}\times\nonumber\\
    &\frac{2(m+1) m N^2 C_{m,N}-4 m^3 N^3 C_{m-1,N}+m^2 (m-1)^2 N^4 C_{m-2,N}}{N^m}R^2\nonumber\\
    &+\mathcal{O}(R^3)\,.
\end{align}
We note that in $C_{j,N}$ the leading order in $N$ comes from the partition $\lambda=(1^j)$, so that in leading order in $N$ we approximate $C_{j,N}\approx C_{j,N}^{(1^j)}=\binom{N}{j}(j!)^2$ in the zeroth and second order. However, in the first order the leading order in $N$ cancels, so that we need to take into account the next leading order in $N$. Therefore, we approximate $C_{j,N}\approx C_{j,N}^{(1^j)} + C_{j,N}^{(1^{j-2},2^1)}=\binom{N}{j}(j!)^2 + \frac{1}{4}\binom{N}{j-1}(j-1)(j!)^2$ in the first order. Then, the intensity correlation function is approximately given by
\begin{align}
    &g^{(m)}(\bm{0}) \approx \frac{(m!)^2 \binom{N}{m}}{N^m}\nonumber\\
    &+ \frac{1}{4} \frac{(m!)^2 \binom{N}{m}m(m-1) N (2N-6m+m^2+8)}{N^{m}(N-m+2)(N-m+1)}R\nonumber\\
	&- \frac{1}{4} \frac{(m!)^2 \binom{N}{m}N^2(N^2+6N+2m-2m^2+4)m(m-1)}{N^{m}(N-m+2)(N-m+1)}R^2\nonumber\\
    &+ \mcal{O}(R^3)\,.
\end{align}
In the limit of $\frac{m!m (m-1)}{2N} \ll 1$, we can further approximate $\frac{m! \binom{N}{m}}{N^m}\approx 1$, so that the correlation function can be approximated by
\begin{align}
    \label{eq:App_GMT_large_N_classical}
	g^{(m)}(\bm{0}) \approx\;& m! + \frac{1}{2} m! m (m-1)R\nonumber\\
	&- \frac{1}{4} m! m (m-1)N^2 R^2 + \mcal{O}(R^3)\,. 
\end{align}
Therefore, the coherence condition for classical oscillators is (in leading order in $N$) the same as for two-level atoms. However, we note the additional factor of $\frac{1}{2}$ and the different sign of the first-order correction ($+$-sign in the present case, $-$-sign in the quantum mechanical case).\\
Now, we come to the finite-$N$ condition for classical oscillators, which is a bit weaker than for two-level atoms. The GMT is given by the leading term in $N$ in the $(1^m)$ partition, whereas the next leading term in $N$ in the $(1^m)$ partition leads to the finite-$N$ condition for two-level atoms. However, in the case of classical oscillators the leading term in $N$ of the partition $(1^{m-2},2^1)$ is of the same order as the second leading term in the $(1^m)$ partition. Therefore, let us first consider again the partition $(1^m)$ and afterwards the partition $(1^{m-2},2^1)$.\\
In the case of the partition $(1^m)$, we find $C_{m,N}^{(1^m)}=\binom{N}{m}(m!)^2$, so that the leading correction is $-\frac{1}{2}m! m (m-1)$ (see quantum mechanical case). In the case of the partition $(1^{m-2},2^1)$, we obtain $C_{m,N}^{(1^{m-2},2^1)}=\frac{1}{4}\binom{N}{m-1}(m-1)(m!)^2$, so that the leading correction is $\frac{1}{4}m! m (m-1)$. The coefficient of the first-order correction ($1/N$) in the classical case is thus $-\frac{1}{4}m! m (m-1)$, i.e., a factor of $2$ smaller than in the quantum mechanical case.

\subsection{Case $m \neq n$}

We consider again the direction $\bm{k}=\bm{0}$ and $m>n$. In the numerator of the correlation function $g^{(m,n)}$, we then have
\begin{align}
    &\braket{[E^*(\bm{0})]^{m-n}[E^*(\bm{0})E(\bm{0})]^n} = \binom{m}{n} (N E^*_\mathrm{coh})^{m-n} \braket{[I(\bm{0})]^n}\nonumber\\
    &= \binom{m}{n} (N E^*_\mathrm{coh})^{m-n} \sum_{j=0}^{n} \binom{n}{j}^2 (N^2 |E_\mathrm{coh}|^2)^{n-j}(|E_\mathrm{incoh}|^2)^j C_{j,N}\,,
\end{align}
so that the normalized correlation function reads
\begin{align}
    g^{(m,n)}(\bm{0}) =\;& \frac{\binom{m}{n} (N E^*_\mathrm{coh})^{m-n}}{N^{\frac{m+n}{2}}(|E_\mathrm{incoh}|^2)^{\frac{m-n}{2}}(1+NR)^{\frac{m+n}{2}}}\times\nonumber\\
    &\sum_{j=0}^{n} \binom{n}{j}^2 (N^2 R)^{n-j} C_{j,N}\,.
\end{align}
Therefore, in zeroth order in $R$ we approximate
\begin{align}
    |g^{(m,n)}(\bm{0})| \approx \frac{\binom{m}{n}N^{m-n} C_{n,N}}{N^{\frac{m+n}{2}}} R^{\frac{m-n}{2}}\,.
\end{align}
The strongest condition is again for $n=m-1$. In addition, we approximate $C_{m-1,N}\approx \binom{N}{m-1}((m-1)!)^2$ in leading order in $N$, so that we obtain
\begin{align}
    \label{eq:App_condSpSm_classical}
    |g^{(m,n)}(\bm{0})| \approx \frac{m N \sqrt{N} \binom{N}{m-1} ((m-1)!)^2}{N^m}\sqrt{R} \approx m! \sqrt{N} \sqrt{R}\,,
\end{align}
which gives the same condition as in the quantum mechanical case.

\section{Comparison of the classical and the quantum mechanical calculation on the examples of $g^{(2)}$ and $g^{(2,1)}$}

\subsection{$g^{(2)}$}

In the quantum mechanical case, we need to calculate
\begin{align}	\sum_{\mu_1,\mu_2,\nu_1,\nu_2=1}^{N} \braket{\hat{\sigma}_{\mu_1}^+ \hat{\sigma}_{\mu_2}^+ \hat{\sigma}_{\nu_2}^- \hat{\sigma}_{\nu_1}^-}\,.
\end{align}
If we write the operators as $\hat{\sigma}_\mu^\pm = \braket{\hat{\sigma}_\mu^\pm} + \delta\hat{\sigma}_\mu^\pm$ we get $16$ different terms, namely
\begin{align}
	&\braket{\hat{\sigma}_{\mu_1}^+ \hat{\sigma}_{\mu_2}^+ \hat{\sigma}_{\nu_2}^- \hat{\sigma}_{\nu_1}^-} =\nonumber\\
    &\braket{\hat{\sigma}_{\mu_1}^+} \braket{\hat{\sigma}_{\mu_2}^+} \braket{\hat{\sigma}_{\nu_2}^-} \braket{\hat{\sigma}_{\nu_1}^-} + \nonumber\\
	&\underbrace{\braket{\delta\hat{\sigma}_{\mu_1}^+} \braket{\hat{\sigma}_{\mu_2}^+} \braket{\hat{\sigma}_{\nu_2}^-} \braket{\hat{\sigma}_{\nu_1}^-}}_{=0} + \underbrace{\braket{\hat{\sigma}_{\mu_1}^+} \braket{\delta\hat{\sigma}_{\mu_2}^+} \braket{\hat{\sigma}_{\nu_2}^-} \braket{\hat{\sigma}_{\nu_1}^-}}_{=0} +\nonumber\\
	&\underbrace{\braket{\hat{\sigma}_{\mu_1}^+} \braket{\hat{\sigma}_{\mu_2}^+} \braket{\delta\hat{\sigma}_{\nu_2}^-} \braket{\hat{\sigma}_{\nu_1}^-}}_{=0} + \underbrace{\braket{\hat{\sigma}_{\mu_1}^+} \braket{\hat{\sigma}_{\mu_2}^+} \braket{\hat{\sigma}_{\nu_2}^-} \braket{\delta\hat{\sigma}_{\nu_1}^-}}_{=0} + \nonumber\\
	&\braket{\delta\hat{\sigma}_{\mu_1}^+ \delta\hat{\sigma}_{\mu_2}^+} \braket{\hat{\sigma}_{\nu_2}^-} \braket{\hat{\sigma}_{\nu_1}^-} + \braket{\delta\hat{\sigma}_{\mu_1}^+ \delta\hat{\sigma}_{\nu_2}^-} \braket{\hat{\sigma}_{\mu_2}^+} \braket{\hat{\sigma}_{\nu_1}^-} +\nonumber\\
    &\braket{\delta\hat{\sigma}_{\mu_1}^+ \delta\hat{\sigma}_{\nu_1}^-} \braket{\hat{\sigma}_{\mu_2}^+} \braket{\hat{\sigma}_{\nu_2}^-} + \braket{\hat{\sigma}_{\mu_1}^+} \braket{\delta\hat{\sigma}_{\mu_2}^+ \delta\hat{\sigma}_{\nu_2}^-} \braket{\hat{\sigma}_{\nu_1}^-} +\nonumber\\
    &\braket{\hat{\sigma}_{\mu_1}^+} \braket{\delta\hat{\sigma}_{\mu_2}^+ \delta\hat{\sigma}_{\nu_1}^-} \braket{\hat{\sigma}_{\nu_2}^-} + \braket{\hat{\sigma}_{\mu_1}^+} \braket{\hat{\sigma}_{\mu_2}^+} \braket{\delta\hat{\sigma}_{\nu_2}^- \delta\hat{\sigma}_{\nu_1}^-} + \nonumber\\
	&\braket{\delta\hat{\sigma}_{\mu_1}^+ \delta\hat{\sigma}_{\mu_2}^+ \delta\hat{\sigma}_{\nu_2}^-} \braket{\hat{\sigma}_{\nu_1}^-} + \braket{\delta\hat{\sigma}_{\mu_1}^+ \delta\hat{\sigma}_{\mu_2}^+ \delta\hat{\sigma}_{\nu_1}^-} \braket{\hat{\sigma}_{\nu_2}^-} + \nonumber\\
	&\braket{\hat{\sigma}_{\mu_1}^+} \braket{\delta\hat{\sigma}_{\mu_2}^+ \delta\hat{\sigma}_{\nu_2}^- \delta\hat{\sigma}_{\nu_1}^-} + \braket{\hat{\sigma}_{\mu_2}^+} \braket{\delta\hat{\sigma}_{\mu_1}^+ \delta\hat{\sigma}_{\nu_2}^- \delta\hat{\sigma}_{\nu_1}^-} + \nonumber\\
    &\braket{\delta\hat{\sigma}_{\mu_1}^+ \delta\hat{\sigma}_{\mu_2}^+ \delta\hat{\sigma}_{\nu_2}^- \delta\hat{\sigma}_{\nu_1}^-}\,.
\end{align}
Now, we have
\begin{align}
	\braket{\delta\hat{\sigma}_{\mu_1}^+ \delta\hat{\sigma}_{\mu_2}^+} &= \begin{cases}
		-\braket{\hat{\sigma}^+}^2 &\mu_1=\mu_2\\
		0 &\mathrm{otherwise}
	\end{cases}\\
	\braket{\delta\hat{\sigma}_{\mu_1}^- \delta\hat{\sigma}_{\mu_2}^-} &= \begin{cases}
		-\braket{\hat{\sigma}^-}^2 &\mu_1=\mu_2\\
		0 &\mathrm{otherwise}
	\end{cases}\\
	\braket{\delta\hat{\sigma}_{\mu_1}^+ \delta\hat{\sigma}_{\mu_2}^-} &= \begin{cases}
		\braket{\delta\hat{\sigma}^+ \delta\hat{\sigma}^-} &\mu_1=\mu_2\\
		0 &\mathrm{otherwise}
	\end{cases}\\
	\braket{\delta\hat{\sigma}_{\mu_1}^+ \delta\hat{\sigma}_{\mu_2}^+ \delta\hat{\sigma}_{\mu_3}^-} &= \begin{cases}
		-2\braket{\hat{\sigma}^+}\braket{\delta\hat{\sigma}^+ \delta\hat{\sigma}^-} &\mu_1=\mu_2=\mu_3\\
		0 &\mathrm{otherwise}
	\end{cases}\\
	\braket{\delta\hat{\sigma}_{\mu_1}^+ \delta\hat{\sigma}_{\mu_2}^- \delta\hat{\sigma}_{\mu_3}^-} &= \begin{cases}
		-2\braket{\hat{\sigma}^-}\braket{\delta\hat{\sigma}^+ \delta\hat{\sigma}^-} &\mu_1=\mu_2=\mu_3\\
		0 &\mathrm{otherwise}
	\end{cases}
\end{align}
and
\begin{widetext}
\begin{align}
   \braket{\delta\hat{\sigma}_{\mu_1}^+ \delta\hat{\sigma}_{\mu_2}^+ \delta\hat{\sigma}_{\mu_3}^- \delta\hat{\sigma}_{\mu_4}^-} &= \begin{cases}
		\braket{\hat{\sigma}^+}^2 \braket{\hat{\sigma}^-}^2 &\mu_1=\mu_2\, \&\, \mu_3=\mu_4\, \&\, \mu_1\neq\mu_3\\
		\braket{\delta\hat{\sigma}^+ \delta\hat{\sigma}^-}^2 &\mu_1=\mu_3\, \&\, \mu_2=\mu_4\, \&\, \mu_1\neq\mu_2\\
		\braket{\delta\hat{\sigma}^+ \delta\hat{\sigma}^-}^2 &\mu_1=\mu_4\, \&\, \mu_2=\mu_3\, \&\, \mu_1\neq\mu_2\\
		\braket{\hat{\sigma}^+} \braket{\hat{\sigma}^-} (\braket{\hat{\sigma}^+} \braket{\hat{\sigma}^-} + 4 \braket{\delta\hat{\sigma}^+ \delta\hat{\sigma}^-}) &\mu_1=\mu_2=\mu_3=\mu_4\\
		0 &\mathrm{otherwise}
	\end{cases}\,.
\end{align}
Therefore, the sum of expectation values evaluates to
\begin{align}
\sum_{\mu_1,\mu_2,\nu_1,\nu_2=1}^{N} \braket{\hat{\sigma}_{\mu_1}^+ \hat{\sigma}_{\mu_2}^+ \hat{\sigma}_{\nu_2}^- \hat{\sigma}_{\nu_1}^-} =\;& N^4 (\braket{\hat{\sigma}^+} \braket{\hat{\sigma}^+})^2 - 2 N^3 (\braket{\hat{\sigma}^+} \braket{\hat{\sigma}^+})^2 + 4 N^3 \braket{\hat{\sigma}^+} \braket{\hat{\sigma}^+} \braket{\delta\hat{\sigma}^+ \delta\hat{\sigma}^-}\nonumber\\
	&- 8 N^2 \braket{\hat{\sigma}^+} \braket{\hat{\sigma}^+} \braket{\delta\hat{\sigma}^+ \delta\hat{\sigma}^-} + N(N-1) (\braket{\hat{\sigma}^+} \braket{\hat{\sigma}^-})^2\nonumber\\
	&+ 2 N(N-1) \braket{\delta\hat{\sigma}^+ \delta\hat{\sigma}^-}^2 + N (\braket{\hat{\sigma}^+} \braket{\hat{\sigma}^-})^2 + 4 N \braket{\hat{\sigma}^+} \braket{\hat{\sigma}^-} \braket{\delta\hat{\sigma}^+ \delta\hat{\sigma}^-}\nonumber\\
	=\;& N^2 (N-1)^2 (\braket{\hat{\sigma}^+} \braket{\hat{\sigma}^+})^2 + 4 N (N-1)^2 \braket{\hat{\sigma}^+} \braket{\hat{\sigma}^+} \braket{\delta\hat{\sigma}^+ \delta\hat{\sigma}^-} + 2 N (N-1) \braket{\delta\hat{\sigma}^+ \delta\hat{\sigma}^-}^2\,.
\end{align}
\end{widetext}
The normalized correlation function thus reads
\begin{align}
    g^{(2)}(\bm{0}) = \frac{1}{N^2 (1+NR)^2} [&2N(N-1)+4N(N-1)^2 R\nonumber\\
    &+N^2 (N-1)^2 R^2]\,.
\end{align}
Performing a Taylor expansion at $R=0$ up to the second order leads to
\begin{align}
    g^{(2)}(\bm{0}) =\;& \frac{2N(N-1)}{N^2} - \frac{4N(N-1)}{N^2}R\nonumber\\
    &-(N-1)(N-7)R^2 + \mathcal{O}(R^3)\,,
\end{align}
which in the large $N$ limit reduces to
\begin{align}
    g^{(2)}(\bm{0}) \approx 2 - 4R - N^2 R^2 + \mathcal{O}(R^3)
\end{align}
in accordance to Eq.~\eqref{eq:App_GMT_gm_Taylor_large_N}.\\
In the classical case, the expectation values $\braket{\delta\hat{\sigma}_{\mu_1}^\pm \delta\hat{\sigma}_{\mu_2}^\pm}$, $\braket{\delta\hat{\sigma}_{\mu_1}^+ \delta\hat{\sigma}_{\mu_2}^+ \delta\hat{\sigma}_{\mu_3}^-}$, and $\braket{\delta\hat{\sigma}_{\mu_1}^+ \delta\hat{\sigma}_{\mu_2}^- \delta\hat{\sigma}_{\mu_3}^-}$ become $0$ due to the fluctuating phase. Note that we simply use the correspondence $\braket{\hat{\sigma}^+}\braket{\hat{\sigma}^-} \leftrightarrow |E_\mathrm{coh}|^2$ and $\braket{\delta\hat{\sigma}_\mu^+\delta\hat{\sigma}_\nu^-} \leftrightarrow |E_\mathrm{incoh}|^2 \left\langle e^{i(\phi_\nu-\phi_\mu)} \right\rangle$. In addition, the expectation value $\braket{\delta\hat{\sigma}_{\mu_1}^+ \delta\hat{\sigma}_{\mu_2}^+ \delta\hat{\sigma}_{\mu_3}^- \delta\hat{\sigma}_{\mu_4}^-}$ reduces to
\begin{widetext}
\begin{align}
    \braket{\delta\hat{\sigma}_{\mu_1}^+ \delta\hat{\sigma}_{\mu_2}^+ \delta\hat{\sigma}_{\mu_3}^- \delta\hat{\sigma}_{\mu_4}^-} &= \begin{cases}
		\braket{\delta\hat{\sigma}^+ \delta\hat{\sigma}^-}^2 &\mu_1=\mu_3\, \&\, \mu_2=\mu_4\, \&\, \mu_1\neq\mu_2\\
		\braket{\delta\hat{\sigma}^+ \delta\hat{\sigma}^-}^2 &\mu_1=\mu_4\, \&\, \mu_2=\mu_3\, \&\, \mu_1\neq\mu_2\\
		\braket{\delta\hat{\sigma}^+ \delta\hat{\sigma}^-}^2 &\mu_1=\mu_2=\mu_3=\mu_4\\
		0 &\mathrm{otherwise}
	\end{cases}\,.    
\end{align}
\end{widetext}
We note that while the first two cases are the same as in the quantum mechanical case (belonging to the $(1^m)$ partition), in the classical case also the third case gives the same as the first two cases (belonging to the $(1^{m-2},2^1)$ partition), which is different in the quantum mechanical case. In the classical case, the correlation function thus reads
\begin{align}
    g^{(2)}(\bm{0}) = \frac{1}{N^2(1+NR)^2}[N(2N-1)+4N^3R+N^4R^2]\,.
\end{align}
Performing a Taylor expansion at $R=0$ up to the second order leads to
\begin{align}
    g^{(2)}(\bm{0}) =\;& \frac{N(2N-1)}{N^2} +2R -(N^2 + 3N)R^2 + \mathcal{O}(R^3)\,,
\end{align}
which in the large $N$ limit reduces to
\begin{align}
    g^{(2)}(\bm{0}) \approx 2 + 2R - N^2 R^2 + \mathcal{O}(R^3)
\end{align}
in accordance to Eq.~\eqref{eq:App_GMT_large_N_classical}.

\subsection{$g^{(2,1)}$}
Similar calculations as in the case of $g^{(2)}$ lead to the following correlation function
\begin{align}
    g^{(2,1)}(\bm{0}) = \frac{1}{N^{\frac{3}{2}}(1+NR)^{\frac{3}{2}}} \Bigg[&2N(N-1) \frac{\braket{\hat{\sigma}^+}}{\sqrt{\braket{\delta\hat{\sigma}^+ \delta\hat{\sigma}^-}}}\nonumber\\
    &+N^2(N-1) \frac{\braket{\hat{\sigma}^+}}{\sqrt{\braket{\delta\hat{\sigma}^+ \delta\hat{\sigma}^-}}} R\Bigg]\,.
\end{align}
In zeroth order in $R$, we then obtain
\begin{align}
    |g^{(2,1)}(\bm{0})| \approx \frac{2N(N-1)}{N^{\frac{3}{2}}} \sqrt{R} \approx 2 \sqrt{N} \sqrt{R}
\end{align}
in accordance to Eq.~\eqref{eq:condSpSm}.\\
In the classical case, the correlation function reads
\begin{align}
    g^{(2,1)}(\bm{0}) = \frac{1}{N^{\frac{3}{2}}(1+NR)^{\frac{3}{2}}} \Bigg[&2N^2 \frac{E_\mathrm{coh}^*}{\sqrt{|E_\mathrm{incoh}|^2}}\nonumber\\
    &+N^3 \frac{E_\mathrm{coh}^*}{\sqrt{|E_\mathrm{incoh}|^2}} R\Bigg]\,.
\end{align}
Thus, in zeroth order in $R$, we obtain
\begin{align}
    |g^{(2,1)}(\bm{0})| \approx \frac{2N^2}{N^{\frac{3}{2}}} \sqrt{R} = 2 \sqrt{N} \sqrt{R}
\end{align}
in accordance to Eq.~\eqref{eq:App_condSpSm_classical}.\\

\end{document}